\documentclass[amsmath,amssymb,nofootinbib,notitlepage,superscriptaddress,twocolumn]{revtex4-1}
\pdfoutput=1

\usepackage{aas_macros,esint,graphicx,subfigure}

\usepackage[colorlinks=true]{hyperref}
\let\originalleft\left
\let\originalright\right
\renewcommand{\left}{\mathopen{}\mathclose\bgroup\originalleft}
\renewcommand{\right}{\aftergroup\egroup\originalright}

\newcommand{\fig}[1]{Fig.\ref{#1}}
\newcommand{\eq}[1]{(\ref{#1})}
\newcommand{\eqs}[1]{(\ref{#1})}

\def\ma{\mathrm}
\def\be{\begin{equation}}
\def\ee{\end{equation}}
\def\ba{\begin{eqnarray}}
\def\ea{\end{eqnarray}}
\def\red{\color{black}}

\def\lf{\left}
\def\rt{\right}
\def\td{\tilde}
\def\pp{\partial}

\begin{document}

\title{Photon sphere and phase transition of $d$-dimensional ($d\ge5$) charged Gauss-Bonnet AdS black holes}

\author{Shan-Zhong Han}
\email{hsz@mail.bnu.edu.cn; Also at Perimeter Institute for Theoretical Physics, Waterloo, Ontario N2L 2Y5, Canada}
\affiliation{Department of Physics, Beijing Normal University, Beijing 100875, China}
\author{Jie Jiang}
\email{jiejiang@mail.bnu.edu.cn}
\affiliation{Department of Physics, Beijing Normal University, Beijing 100875, China}
\author{Ming Zhang}
\email{mingzhang@mail.bnu.edu.cn; Also at Department of Physics and Astronomy, University of Waterloo, On tario N2L 3G1,Canada}
\affiliation{Department of Physics, Beijing Normal University, Beijing 100875, China}
\affiliation{College of Physics and Communication Electronics, Jiangxi Normal University, Nanchang 330022, China}
\author{Wen-Biao Liu}
\email{wbliu@bnu.edu.cn}
\affiliation{Department of Physics, Beijing Normal University, Beijing 100875, China}

\begin{abstract}
Motivated by recent work, nonmonotonic behaviours of photon sphere radius can be used to reflect black hole phase transition for Reissner-Nordstr$\ddot{o}$m-AdS (RN-AdS) black holes, we study the case of five-dimensional charged Gauss-Bonnet-AdS (GB-AdS) black holes in the reduced parameter space. We find that the nonmonotonic behaviours of photon sphere radius still exist. Using the coexistence line calculated from $P-V$ plane, we capture the photon sphere radius of saturated small and large black holes (the boundary of the coexistence phase), then illustrate the reduced coexistence region. The results show that, reduced coexistence region decreases with charge $Q$ but increases with Gauss-Bonnet coefficient  $\alpha$.  When the charge vanishes, reduced coexistence region doesn't vary with Gauss-Bonnet coefficient $\alpha$ any more. In this case, the Gauss-Bonnet coefficient $\alpha$ plays the same role as the charge of five-dimensional RN-AdS black holes. Also, the situation of higher dimension is studied in the end.

\end{abstract}

\maketitle

\section{Introduction}

In the past decades, black hole thermodynamics has been an area of intense investigation \cite{1973PhRvD...7.2333B,1974Natur.248...30H,1977Natur.266..333R,Blandford:1977ds,1976PhRvD..13..198P,1977PhRvD..15.2738G,1982CMaPh..87..577H,1995PhRvL..75.1260J,1999PhRvL..82.4971C,2003PhRvD..68d6005K,2015PhRvD..91l4033X,2002RvMP...74..825B,1999AIPC..484...51M,1998PhLB..428..105G,1998AdTMP...2..253W,2015PhRvD..91f4046C,2015PhRvD..92h1501Z,2018PhLB..787...64W,2018IJTP...57.3429H}. A series of work shows that there exist rich phase structures and many different thermodynamic properties in black hole spacetime. One of the most interesting things is that small black holes are found to be thermodynamically unstable while large ones are really stable in anti-der Sitter (AdS) space. There is a minimum temperature for AdS black holes, no black hole solution can be found below this temperature. Then Hawking-Page phase transiton was proposed \cite{1982CMaPh..87..577H}. It is from this seminal start that various research sprung up \cite{2000PhRvD..62b4027H,2001CMaPh.217..595D,2004PhRvL..92n1301C,2005PhRvL..94k1601K,2006PhRvL..97k1601P,Eisert:2008ur,2011PhRvL.107j1602T,2013JHEP...02..062A,2014okml.book..389W,2010CQGra..27w5014K,2010CQGra..27w5014K,2011CQGra..28s5022K,2013JHEP...09..005C,2014Galax...2...89A,2013PhRvD..88h4045H}.

Regarding cosmological constant as a dynamical pressure and its conjugate quantity as a thermodynamic volume, Kubiznak and Mann at the first time showed that, for a charged AdS black hole, the system indeed has a first-order small-large black hole phase transition which in many aspects resembles the liquid-gas phase transition occurring in fluids \cite{2012JHEP...07..033K,2012JHEP...11..110G}. Although taking the cosmological constant as a thermodynamic variable is not a consensual technique amongst the high energy physics community, there are still at least three advantages inspiring scientists. i) Their investigation of 4-dimensional charged AdS black hole thermodynamics indicates that the black hole system is strikingly similar to the Van der Waals fluid. It seems more meaningful in the fundamental theories \cite{1996PhRvL..77.4992G,1995PhRvD..52.4569C}. ii)Moreover, only when the cosmological constant is taken into account can the thermodynamics of black holes be consistent with Smarr relation \cite{2009CQGra..26s5011K,2005GReGr..37..643B,yi2010energy}. iii) Since cosmological constant $\Lambda$ corresponds to pressure, using geometric units $G_{N}=\hbar=c=k=1$, one identifies the dynamical pressure with $P=-\frac{1}{8\pi}\Lambda=\frac{3}{8\pi}\frac{1}{l^2}$ for a 4-dimensional asymptotically AdS black hole. It's natural to consider its conjugate quantity as volume and to conject it satisfying the reverse isoperimetric inequality \cite{2009CQGra..26s5011K}. Therefore, many people have been attracted on the black hole thermodynamics in the extended phase space. Various novel thermodynamic properties have been discovered, such as the triple point, reentrant phase transition and superfluid black hole phase etc \cite{2013PhRvD..88j1502A,2014CQGra..31d2001A,2014JHEP...09..080F,2014PhRvD..90d4057W,2014CQGra..31x2001D,2015JHEP...07..077H,2016PhRvD..93h4015W,2017PhRvL.118b1301H,2017arXiv170704101Z,2015JHEP...11..157H,2016CQGra..33w5007H,1998mcsp.book.....R,2017PhRvD..95b1501H,2017PhLB..765..154M,2015PhRvD..92j4011C,2017CQGra..34f3001K,2017GReGr..49...57L,2016EPJC...76..571H}. At the same time, there are many attempts to find  an observational path to reveal the thermodynamic phase transition of the black hole \cite{2014JHEP...09..179L,2016JHEP...04..142M,2016EPJC...76..676C,2017EPJC...77..365Z,2017EPJC...77...27P,2017arXiv171207812L}. For example, Ref. \cite{2014JHEP...09..179L} illustrated that, with the value of the horizon radius increasing, the slopes of the quasinormal frequency change drastically different in the small and large black holes. This provides the expectation to find observable signature of black hole phase transitions. 

Recently, the relationship between the unstable circular photon orbit and thermodynamic phase transition in Einstein-AdS spacetime was studied \cite{2018PhRvD..97j4027W,2019PhRvD..99d4013W,2019PhRvD..99f5016Z}. They found that, no matter {\red{$d$}}-dimensional charged RN-AdS black holes or the rotating Kerr-AdS black holes, the radius of unstable circular photon orbits can have oscillating behaviours below the critical point. They present a significant conjecture that thermodynamic phase transition information can be revealed by the behaviours of the radius of unstable circular photon orbit. Their conjecture leads us to think about a question, can this be applied to the modified Einstein theory? e.g. Gauss-Bonnet AdS black holes. On the other hand, there are many recent works discussing the microscopic structure of black holes \cite{Wei:2015iwa,Wei:2019uqg}. As they mentioned, the equation of state does not apply in the coexistence region. Thus, it would provide deep understanding for black hole thermodynamics if we can experimentally capture the boundary of this coexistence region$--$saturated small/large black hole. Based on the above two points, we investigate the relationship between small black hole-large black hole (SBH-LBH) phase transition and photon sphere radius in Einstein-Gauss-Bonnet-AdS spacetime. We hope this work may provide some inspiration for astronomical observation to determine the coexistence phase of black holes.

In this paper, firstly, we derive the photon sphere radius which is dependent on pressure {\red{$P$}}. Then, according to thermodynamics theory, we investigate the relationship between the thermodynamic temperature and the photon sphere radius in the reduced {\red{parameter}} space. After that, we plot the coexistence line with the help of the equal area law in $P-V$ plane. Using it, the phase transition temperature for arbitrary pressure is obtained. The photon sphere radius of saturated small/large black holes can be determined. Based on these, it will be discussed how {\red{reduced}} photon sphere radius of saturated small/large black holes varies with the charge and Gauss-Bonnet coefficient. We also discuss the impact of dimension in the end.

In Sec.~\ref{sec2}, we review the definition of the photon sphere. In Sec.~\ref{sec3}, we discuss the thermodynamics of $d (d\ge5)$-dimensional charged GB-AdS black hole. In Sec.~\ref{sec4}, we study the relationship between the photon sphere radius and SBH-LBH phase transition for the case of $d=5$ and the case of $d>5$ respectively. Finally, the discussions and conclusions are given in Sec.~\ref{sec5}.

\section{Photon sphere}\label{sec2}

For a {\red{$d$}}-dimensional spherically symmetric black hole, the line element is
         \be
              \ma{d}s^2=-f(r)\ma{d}t^2+\frac{1}{f(r)}\ma{d}r^2+r^2\ma{d}\Omega^2_{d-2},
         \ee
where {\red{$\ma{d}\Omega^2_{d-2}$}} is the metric element on the unit $(d-2)$-dimensional sphere, for which the angular coordinates are $\theta_i \in [0,\pi]$ $(i=1,...,d-3)$ and $\phi \in [0,2\pi$]. Considering a photon moving in the equatorial plane where the angular coordinates $\theta_i=\pi/2$ for $i=1,2,...,d-3$, the Lagrangian is
         \be\label{L}
              \mathcal{L}=\frac{1}{2}\lf(-f(r)\lf(\frac{\ma{d}t}{\ma{d}\lambda}\rt)^2+\frac{1}{f(r)}\lf(\frac{\ma{d}r}{\ma{d}\lambda}\rt)^2+r^2\lf(\frac{\ma{d}\phi}{\ma{d}\lambda}\rt)^2\rt),
         \ee
where $\lambda$ is an affine parameter. The metric has two Killing vectors $\pp/\pp t$ and $\pp/\pp\phi$, so there exist two conserved quantities, energy $E$ at infinity and orbital angular momentum $L$ of the photon as
         \ba\label{2-Conserved}
              \begin{aligned}
                                    f(r) \frac{\ma{d}t}{\ma{d}\lambda}=E,\\
                                    r^2 \frac{\ma{d}\phi}{\ma{d}\lambda}=L.
              \end{aligned}
         \ea
Since the photon is moving along null geodesic, we also have
         \be\label{2-Null}
              -f(r)\lf(\frac{\ma{d}t}{\ma{d}\lambda}\rt)^2+\frac{1}{f(r)}\lf(\frac{\ma{d}r}{\ma{d}\lambda}\rt)^2+r^2\lf(\frac{\ma{d}\phi}{\ma{d}\lambda}\rt)^2=0.
         \ee
Combining Eq.\eqs{2-Conserved} and Eq.\eqs{2-Null}, the radial motion equation can be derived as
         \be\label{2-Radial}
               \lf(\frac{\ma{d}r}{\ma{d}\lambda}\rt)^2+\frac{f(r)}{r^2}L^2=E^2.
         \ee
It can be rewritten as
         \be
              \lf(\frac{\ma{d}r}{\ma{d}\xi}\rt)^2+\frac{f(r)}{r^2}=\frac{E^2}{L^2},
         \ee
if we use a new affine parameter $\xi=L\lambda$. Setting $\chi=\frac{L}{E}$, we have
         \be
              \lf(\frac{\ma{d}r}{\ma{d}\xi}\rt)^2+V_{\mathrm{eff}}(r)=\chi^{-2}.
         \ee
Here, we define the effective potential as
         \be
              V_{\mathrm{eff}}(r)=f(r)/r^2.
         \ee
         
\begin{figure}[htp]
	\centering
	\includegraphics[width=\columnwidth]{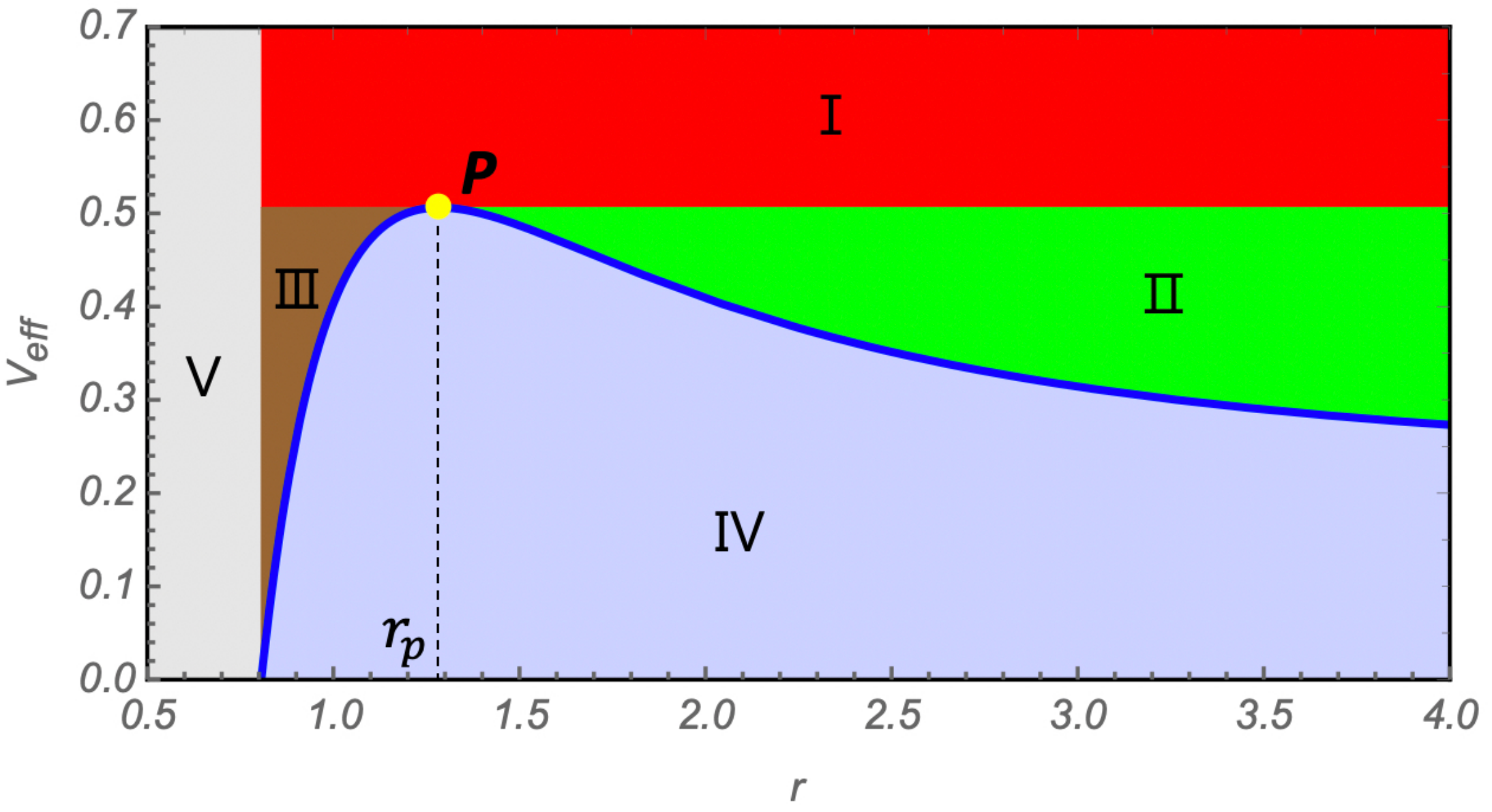}
	\caption{{\red{The effective potential $V_{\ma{eff}}(r)$ of a 5-dimensional charged Gauss-Bonnet AdS black hole}} denoted by blue curve. The impact parameter of the photons moving in each region satisfys the following conditions respectively: \uppercase\expandafter{\romannumeral1}: $\chi<{\red{\sqrt{V_{\ma{eff}}(P)}}}$; \uppercase\expandafter{\romannumeral2}: $\chi>{\red{\sqrt{V_{\ma{eff}}(P)}}}$; \uppercase\expandafter{\romannumeral3}: $\chi>{\red{\sqrt{V_{\ma{eff}}(P)}}}$. For photons in region \uppercase\expandafter{\romannumeral3}, no matter they are ingoing or outgoing, they will all fall into the black hole in the end. Region \uppercase\expandafter{\romannumeral4} denotes the potential barrier, region \uppercase\expandafter{\romannumeral5} denotes the interior of the black hole, yellow point $P$ corresponds to photon sphere {\red{on which the photon has the maximum effective potential $V_{\ma{eff}}(P)$}}. Here, we set {\red{$M=1, Q=0.02, \alpha=0.1, P=0.05$}}.}
	\label{2-schA}
\end{figure}

\fig{2-schA} shows the effective potential of a {\red{5-dimensional charged Gauss-Bonnet AdS}} black hole (blue line). {\red{In this case, $f(r)$ reads

         \be
         f=1+\frac{r^{2}}{2 \alpha}(1-\sqrt{1+\frac{32 M \alpha}{3 \pi r^{4}}-\frac{16 \pi \alpha P}{3}-\frac{64\alpha Q^{2}}{3 r^{6}}}).
         \ee
         
Apparently, to reach {\red{$r=r^* (r^*>r_p)$}} for ingoing photons from infinity, the impact parameter $\chi$ needs to satisfy
         \be
              \chi^{-2}\ge V_{\ma{eff}}(r^*).
         \ee}}
       
For simplicity, we {\red{only}} consider the photons moving inward from infinity (more details can be found in Refs. \cite{1959RSPSA.249..180D,1961RSPSA.263...39D,2001JMP....42..818C,2003PhRvD..67l4017D,2010PhRvD..81j4039D,1973grav.book.....M,2017mcp..book.....T,2008CQGra..25x5009G}). i) Photons with $\chi<{\red{\sqrt{V_{\ma{eff}}(P)}}}$ will move through the horizon and eventually fall into the black hole. ii) Photons with $\chi>{\red{\sqrt{V_{\ma{eff}}(P)}}}$ will reach the potential barrier (periastron) and then escape to infinity; specially, for those photons with ${\red{(\chi-\sqrt{V_{\ma{eff}}(P)}})\to 0^+}$, they will circle the black hole many times at $r$$\approx${\red{$r_p$}} (unstable circular orbit) before escaping to infinity. iii)Photons with  $\chi={\red{\sqrt{V_{\ma{eff}}(P)}}}$ circling the black hole (at {\red{$r=r_p$}}) will deviate from their orbits as long as there is a little perturbation. Actually, the surface $r={\red{r_p}}$ is the boundary between escaping to infinity and falling into the black hole, we call it photon sphere\footnote{For a rotating Kerr black hole, it corresponds to an unstable circular orbit.}.  The radius of photon sphere can be determined by
         \be\label{2-Veff}
              \frac{\partial V_{\ma{eff}}}{\partial r}\bigg|_{r_{p}}=0,\ \ \ \ \ \ \  \frac{\partial^2V_{\ma{eff}}}{\partial r^2}\bigg|_{r_{p}}<0.
         \ee

\section{Thermodynamics of charged Gauss-Bonnet AdS black holes}\label{sec3}

Generally, for GB-AdS spacetime, there are three cases: $k=-1$, $k=0$, $k=+1$. $k=-1$ corresponds to constant negative curvature on a maximally symmetric space $\Sigma$; $k=0$ corresponds to no curvature on $\Sigma$; $k=+1$ corresponds to positive curvature on $\Sigma$. Since according to Ref.\cite{2013JHEP...09..005C}, in the case of $k=1$ and $d\ge5$, there exists small-large black hole phase transition, we will focus on this situation in the following.

The metric of a {\red{$d$}}-dimensional charged GB-AdS black hole with a negative cosmological constant $\Lambda=-(d-1)(d-2)/2l^2$ is
         \be
         \ma{d}s^2=-f(r)\ma{d}t^2+\frac{1}{f(r)}\ma{d}r^2+r^2\ma{d}\Omega^2_{d-2}.
         \ee
The metric function is\cite{WILTSHIRE198636,Wang:2019urm,2002PhRvD..65h4014C,2013PhRvD..87d4014W}
         \begin{align}
         f(r)&=1+\frac{r^{2}}{2 \alpha}\Bigg(1- \nonumber \\ & \sqrt{1+\frac{2 \alpha}{d-2}\left(\frac{32 \pi M}{A r^{d-1}}-\frac{64\pi Q^{2}}{(d-3) r^{2 d-4}}-\frac{32 \pi P}{d-1}\right)}\Bigg),
         \label{3-GSB-f(r)}
         \end{align}
where $\alpha=2\alpha_\text{GB}$, $\alpha_\text{GB}$ is the Gauss-Bonnet coefficient with dimension $[\text{length}]^2$, $M$ denotes the black hole mass, $Q$ corresponds to  the charge of the black hole and $P =-\frac{\Lambda}{8\pi}=\frac{3}{4\pi l^2}$, $A$ is the area of a unit $(d-2)$-dimensional sphere as
         \be
              A=\frac{2 \pi^{(d-1) / 2}}{\Gamma[(d-1) / 2]}.
         \ee
In the following, we will replace the variable $M$ with the horizon $r_h$ which is the largest real root of the equation $f(r_h)=0$.
         \begin{align}
         M=&\frac{A}{16 \pi}\left((d-2) r_{h}^{d-3}+\frac{16 P \pi r_{h}^{d-1}}{d-1}+\frac{8Q^{2}}{(d-3) r_{h}^{d-3}}\right)+ \nonumber\\ & \frac{A}{16 \pi}(d-2) r_{h}^{d-5} \alpha.
         \label{3-GSB-M}
         \end{align}
The black hole temperature is given by

         \be\label{3-GSB-T}
         T=\frac{(d-3) r_{h}^{2}+(d-5) \alpha}{4 \pi r_{h}\left(2 \alpha+r_{h}^{2}\right)}+\frac{4 P r_{h}^{3}-8 Q^{2} r_{h}^{7-2 d}}{(d-2)\left(2 \alpha+r_{h}^{2}\right)}.
         \ee

From Eq.$\eqs{3-GSB-T}$, the equation of state of the black hole can be given as
         \begin{align}
         P=\frac{(d-2)\left(2 \alpha+r_{h}^{2}\right)}{4 r_{h}^{3}} &T+2Q^{2} r_{h}^{4-2 d}- \nonumber\\ & \frac{(d-2)\left((d-5) \alpha+(d-3) r_{h}^{2}\right)}{16 \pi r_{h}^{4}}.
         \end{align}

Then the volume and specific volume are identified as \cite{2013JHEP...09..005C}
         \be\label{Volume}
         V=\frac{A r_{h}^{d-1}}{d-1}, \quad \quad \quad \quad v=\frac{4 r_{h}}{d-2}.
         \ee
One can see that the specific volume $v$ depends linearly on the horizon radius $r_h$.  Without loss of generality, we will replace $v$ with $r_{h}$. The critical point in this case can be obtained from
         \be\label{3-GSB-critical}
         \frac{\pp P}{\pp r_{h}}=0,\quad \quad \quad \quad \frac{\pp^2 P}{\pp^2 r_{h}}=0.
         \ee   

\section{Small-large black hole phase transition and photon sphere}\label{sec4}

\subsection{{\red{$d=5$}} dimensional charged GB-AdS black holes }

\begin{figure}[t]
        \centering
        \subfigure[]{\label{coexistence}\includegraphics[width=\columnwidth]{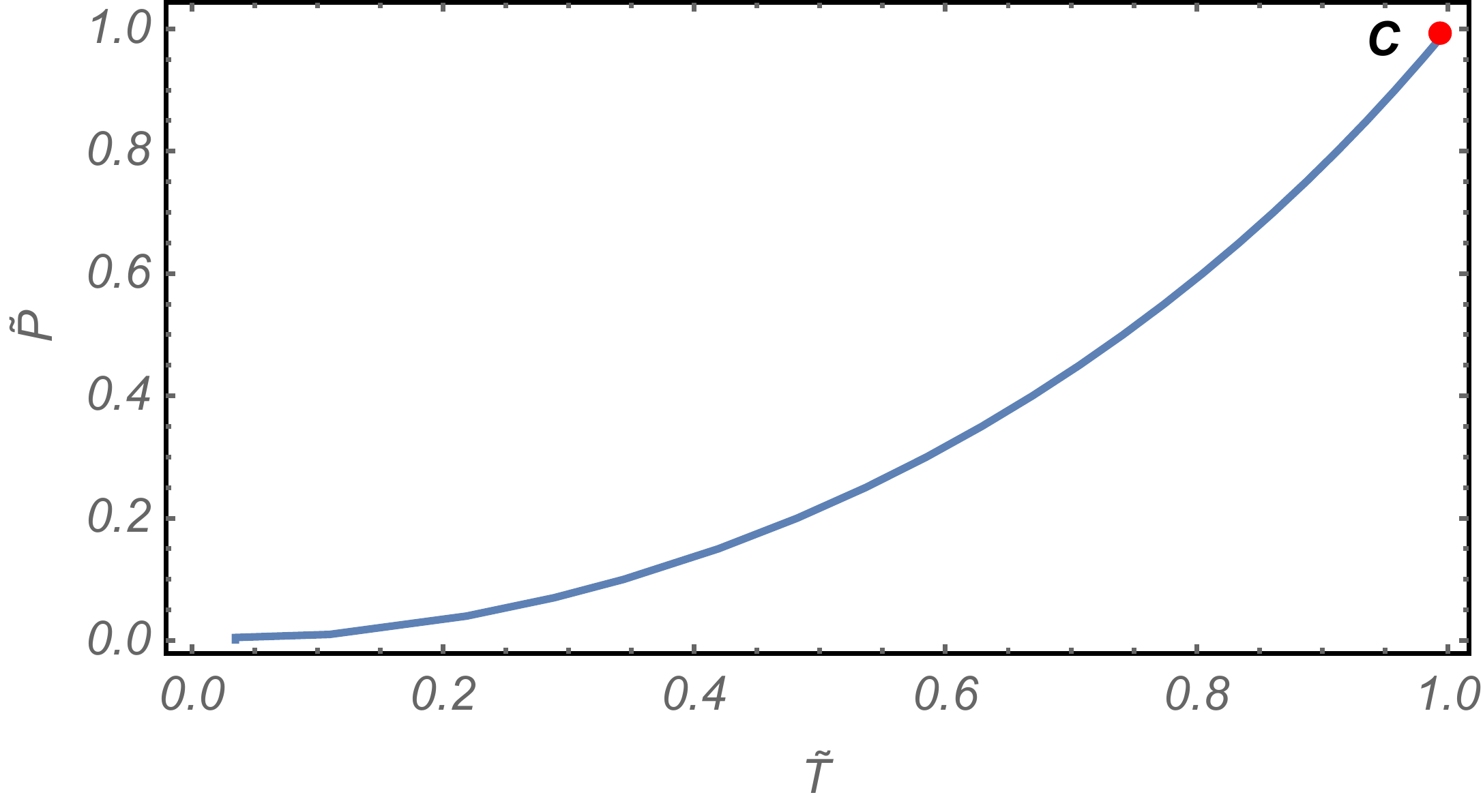}}\\
        \subfigure[]{\label{Ttrpt}\includegraphics[width=\columnwidth]{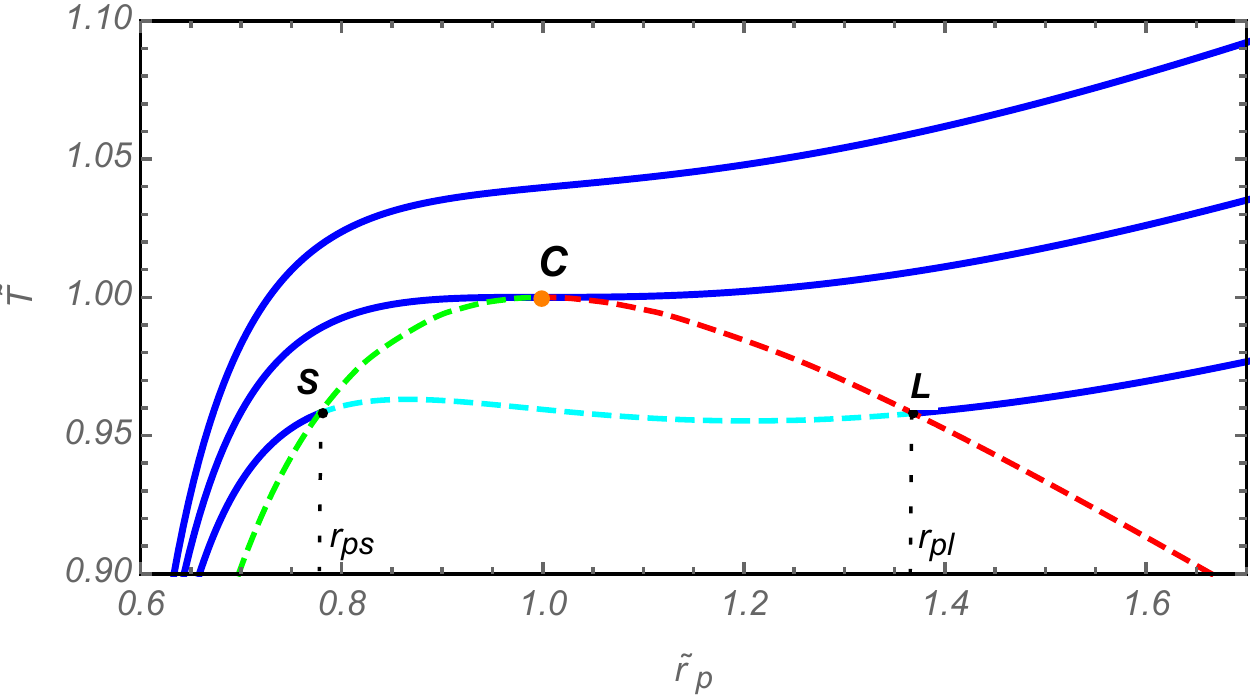}} 
        \caption{(a) coexistence curve for 5-dimensional charged GB-AdS black holes; (b) isobaric curves (blue solid line) with $\tilde{P} = 0.9, 1.0$ and $1.1$ from bottom to top in $\tilde{T}-\tilde{r_{p}}$ diagram. Cyan dashed curve denotes the state calculated from the equation of state. Note that in this coexistence region, the equation of sate is invalid. Green (left) and Red (right) dashed curves correspond to saturated small and large black holes. Black dot $S$ and $L$ correspond to saturated small and large black hole for constant pressure $\tilde{P}=0.9$. Orange dot $C$ denotes critical point. $r_{ps}$ and $r_{pl}$ denote the photon sphere radius of small and large black holes. Here, we set $Q=1,\alpha=0.2$.}
        \label{fig:inspiral}
\end{figure}

In this case, the equation of state is
         \be\label{5P}
         P=\frac{3 \left(2 \alpha+r_{h}^{2}\right)}{4 r_{h}^{3}}T-\frac{3}{8 \pi r_{h}^{2}}+\frac{2Q^{2}}{r_{h}^{6}}.
         \ee
Thinking of Eq.\eqs{3-GSB-critical}, we know that the critical point should satisfy
         \ba\label{5critical}
         \begin{aligned}
         16\pi Q^2-r_{hc}^4+6\pi\alpha r_{hc}^3T_c+\pi r_{hc}^5T_c&=0,\\
        112\pi Q^2-3r_{hc}^4+24\pi\alpha r_{hc}^3T_c+2\pi r_{hc}^5T_c&=0.
         \end{aligned}
         \ea
{\red{Based on the critical point we derived,}} we shall present the thermodynamic quantities in the reduced parameter space in the following. A reduced quantity is defined as $\tilde{X}=X/X_c$, where the index c denotes a quantity that takes the value at the critical point of thermodynamic phase transition. 

Although we are able to calculate the critical point using above equations, however, equal area law may not be valid in $P-v$ plane\cite{2015EPJC...75...71B,2015EPJC...75..419L}. So we derive the coexistence line in $P-V$ plane. From Eq.\eq{Volume}, we have
         \be\label{5P}
         P=\frac{3}{4 \sqrt{2} {W}^{1 / 4}}\left(1+\frac{\alpha}{{W}^{1 / 2}}\right) {T}-\frac{3}{16 \pi {W}^{1 / 2}}+\frac{{Q}^{2}}{4{W}^{3 / 2}},
         \ee
where
         \be
         W=\frac{V}{A}.
         \ee
Without loss of generality, we regard $W$ as the volume. Using the equal area rule
         \be
         \int_{W_{1}}^{W_{2}} {P}\left(\alpha, {Q}, {T}^{*}, {W}\right) {\ma{d}} {W}={P}^{*}({W} 2-{W} 1),
         \ee
and with the help of Eq.\eq{5P}, we plot the coexistence line in \fig{coexistence}.

On the other hand, from Eq.\eqs{2-Veff}, the radius of photon sphere denoted by $r_{p}$ can de derived from
         \begin{align}\label{5ps}
         -96 \sqrt{3} \pi {Q}^{2} {r}_{{h}}^{2}+4& \sqrt{3}\left(16 \pi {Q}^{2}+3 \alpha {r}_{{h}}^{2}+3 {r}_{{h}}^{4}+4 {P} \pi r_{{h}}^{6}\right) {r}_{{p}}^{2}\nonumber\\&-6r_{h}^{2} r_{p}^{4}\sqrt{3-16\pi \alpha P+X/r_{p}^{6}}=0, 
         \end{align}
where
         \begin{align}
         X=&4 \alpha\left(\left(3 \alpha+16 \pi Q^{2} / r_{h}^{2}+3 r_{h}^{2}+4 P \pi r_{h}^{4}\right) r_{p}^{2}-16 \pi Q^{2}\right).\nonumber\\&
         \end{align}

Together with Eq.\eqs{5P}, it's clearly that the temperature $T$ is just a function of photon sphere radius $r_p$ and pressure $P$ for fixed charged $Q$ and Gauss-Bonnet coefficient $\alpha$. \fig{Ttrpt} illustrates the isobaric lines (blue solid lines) between the temperature $T$ and photon sphere radius $r_p$ in the reduced space. As we can see, there exist nonmonotonic behaviours of photon sphere radius below the critical point which is analogous to the Van der Waals system. It means that photon sphere radius can not only  be regarded as a characteristic variable in Einstein-AdS spacetime\cite{2018PhRvD..97j4027W,2019PhRvD..99d4013W}, but also in Einstein-Gauss-Bonnet-AdS spacetime.

As shown in \fig{Ttrpt}, using coexistence line, given a pressure, there are two special photon spheres which correspond to saturated small black hole (point $S$) and saturated large black hole (point $L$), respectively. For simplicity, we shall call them small photon sphere and large photon sphere in this paper. Between them, it's {\red{reduced}} coexistence region {\red{$\Delta \td{r}_{p} (=\td{r}_{pl}-\td{r}_{ps})$}}. As we all know, the equation of state (cyan dashed line) is invalid in this coexistence region. In order to have a deep understanding for black hole thermodynamics and black hole microstruture, it's worth paying more attention to this region. We illustrate the range of the {\red{reduced}} coexistence region in \fig{Ttrpt} (between the green dashed line and red dashed line). As we can see, small photon sphere coincides with the large one at critical point $C$ where it corresponds to second order black hole phase transition.

\begin{figure*}[htbp]
	\centering
        \subfigure[]{\label{smallQ5}\includegraphics[width=5.55cm]{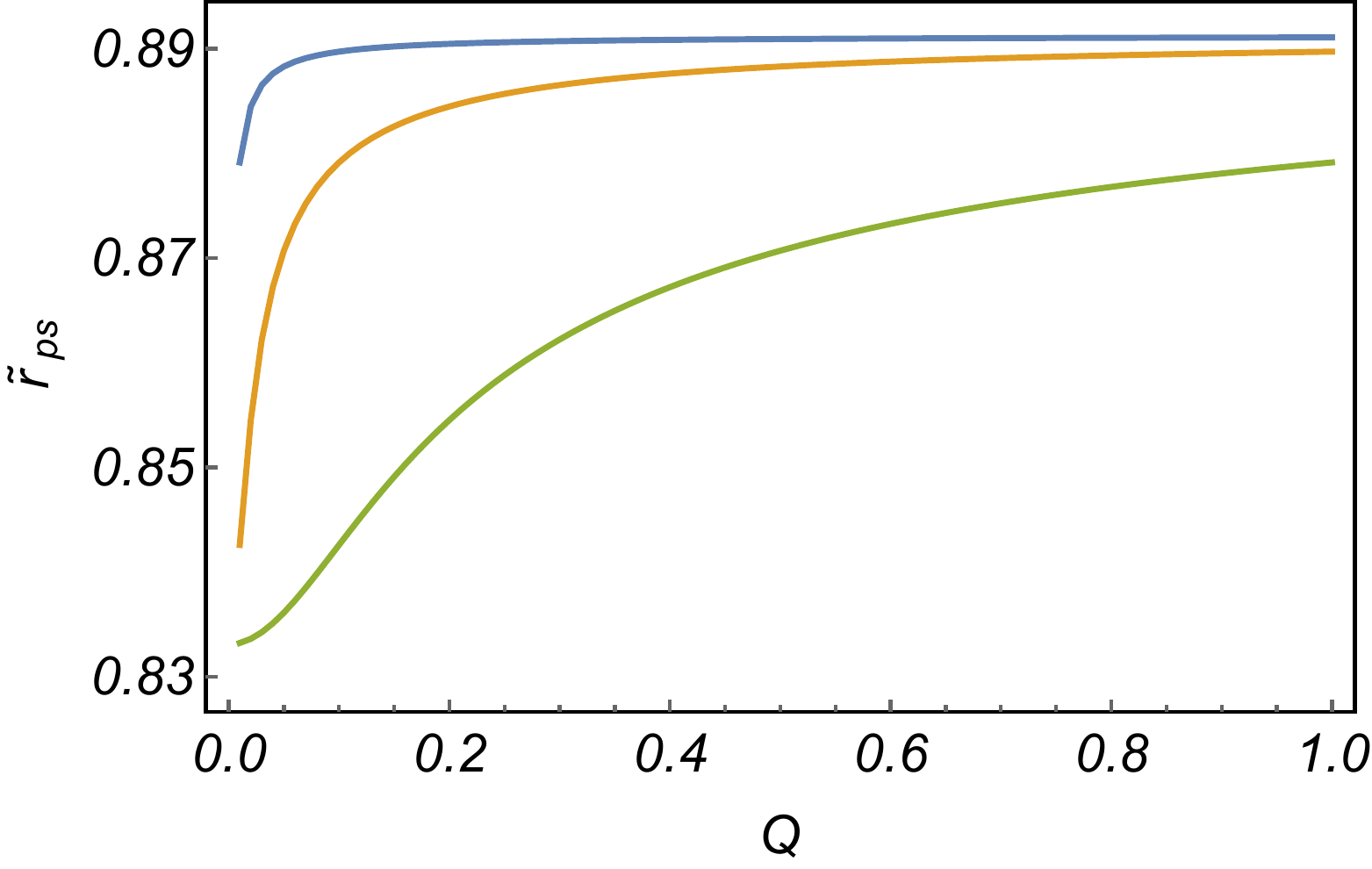}}
        \subfigure[]{\label{largeQ5}\includegraphics[width=5.55cm]{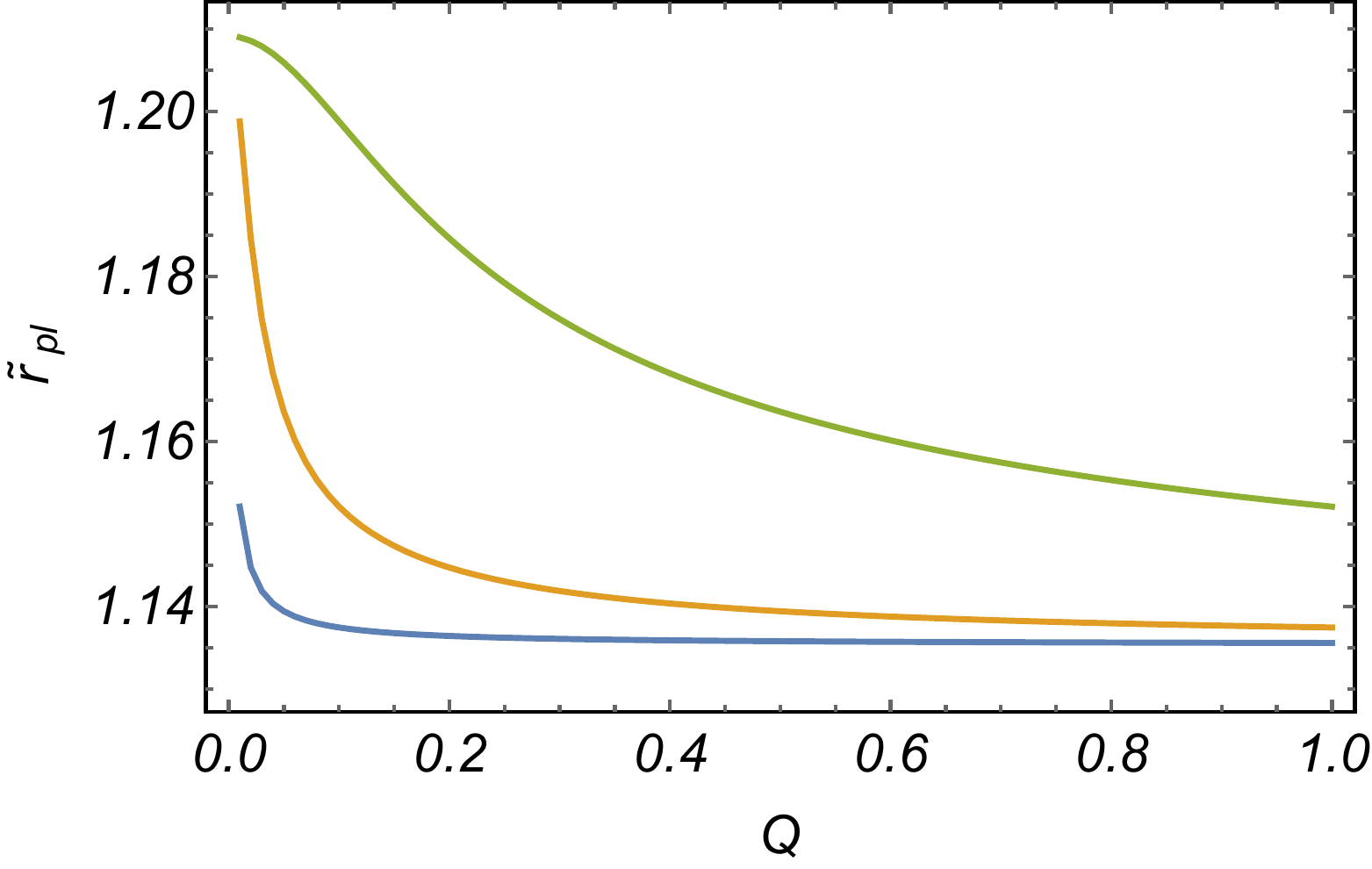}}
        \subfigure[]{\label{coQ5}\includegraphics[width=5.55cm]{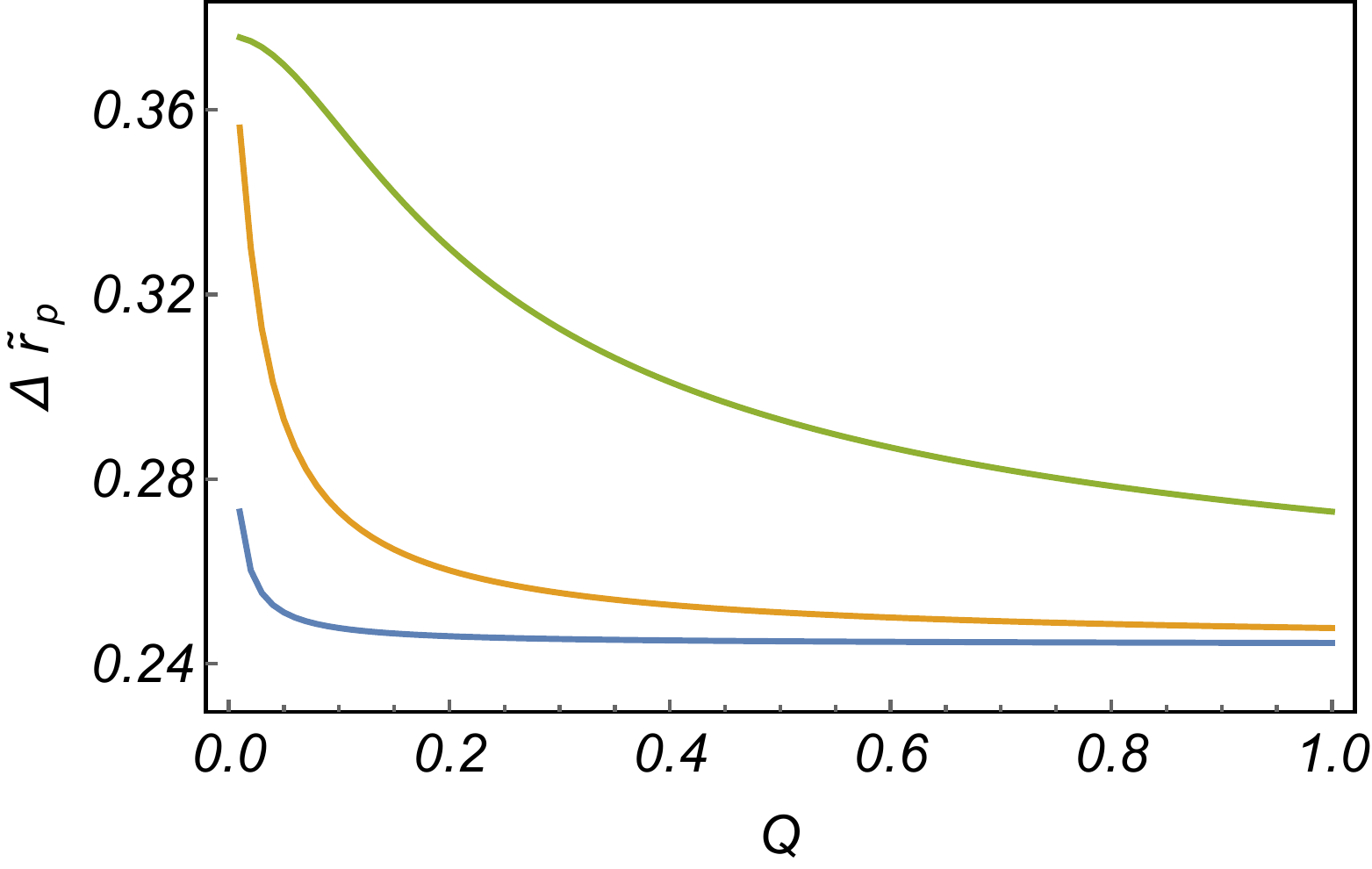}} 
	\caption{Behaviours of the {\red{reduced}} small/large photon sphere radius as a function of {\red{$Q$ for $5$-dimensional Gauss-Bonnet AdS black holes with coefficient $\alpha = 0.01$(blue line), $0.1$(orange line) and $1$(green line) at}} $\tilde{P}=0.98$ : (a) $\td{r}_{ps}$, (b) $\td{r}_{pl}$, (c) $\Delta \td{r}_{p}$.}
	\label{5Q}
\end{figure*}

\begin{figure*}[htbp]
        \centering
        \subfigure[]{\label{smalla5}\includegraphics[width=5.55cm]{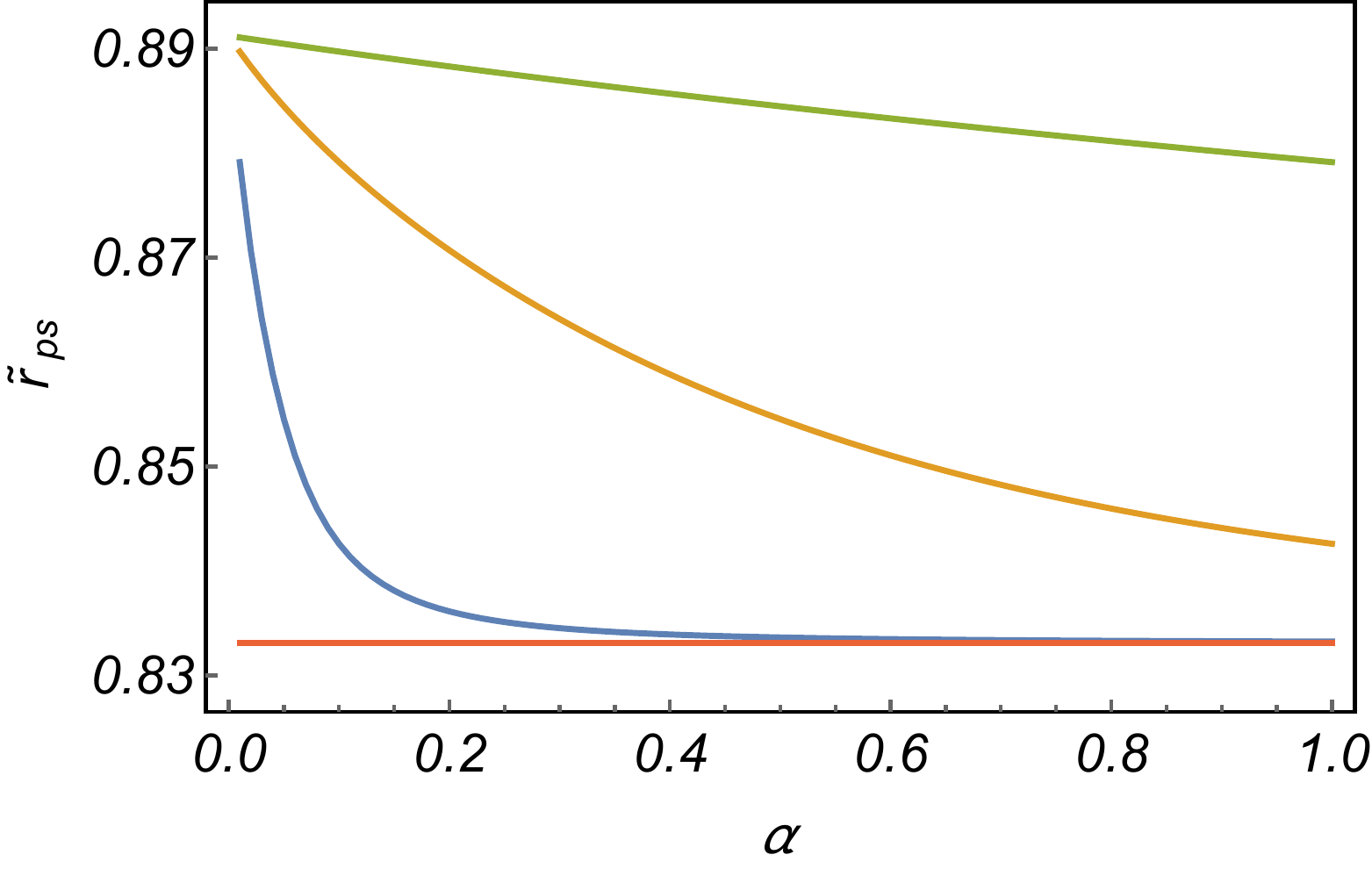}}
        \subfigure[]{\label{largea5}\includegraphics[width=5.55cm]{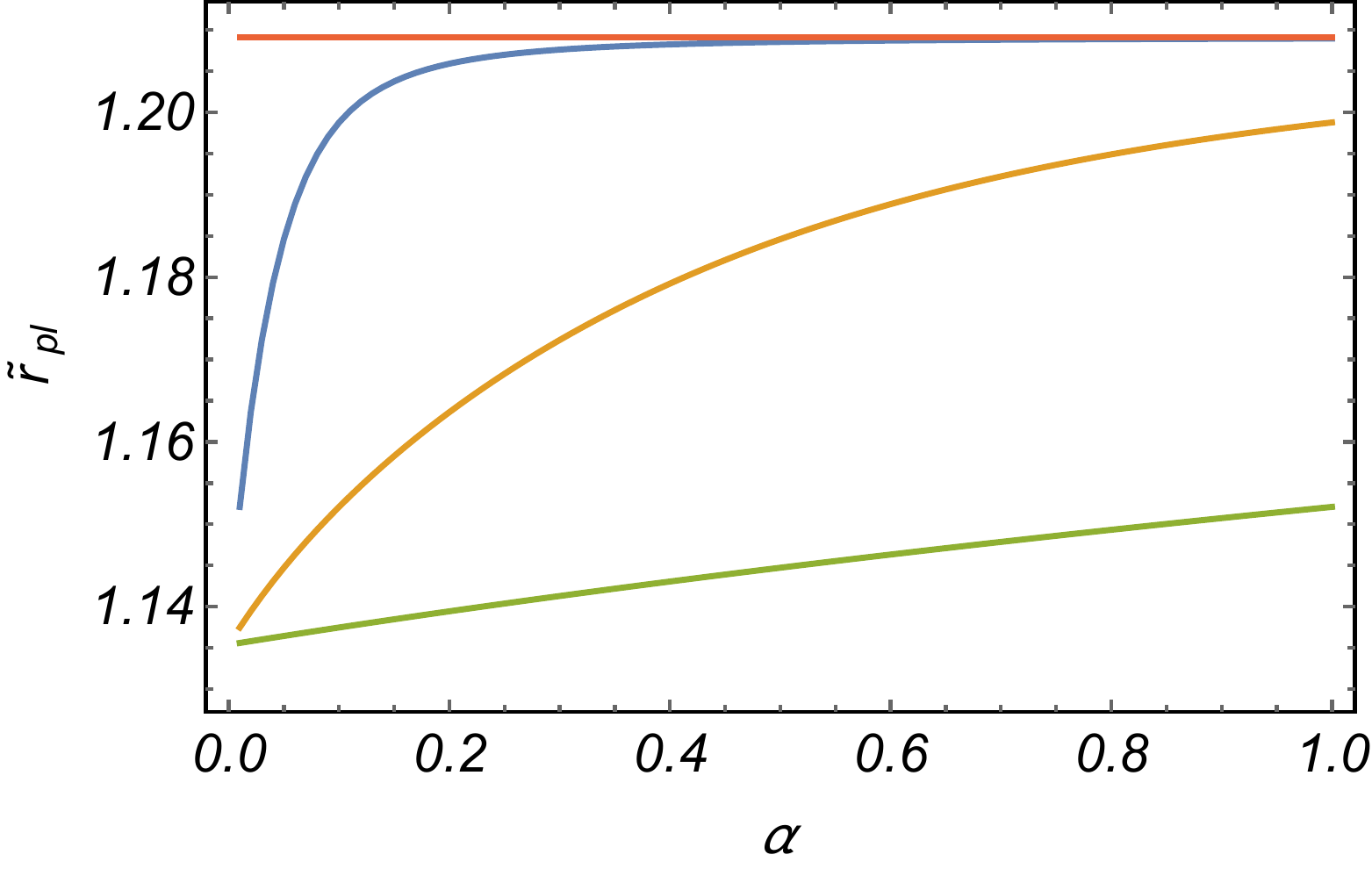}}
        \subfigure[]{\label{coa5}\includegraphics[width=5.55cm]{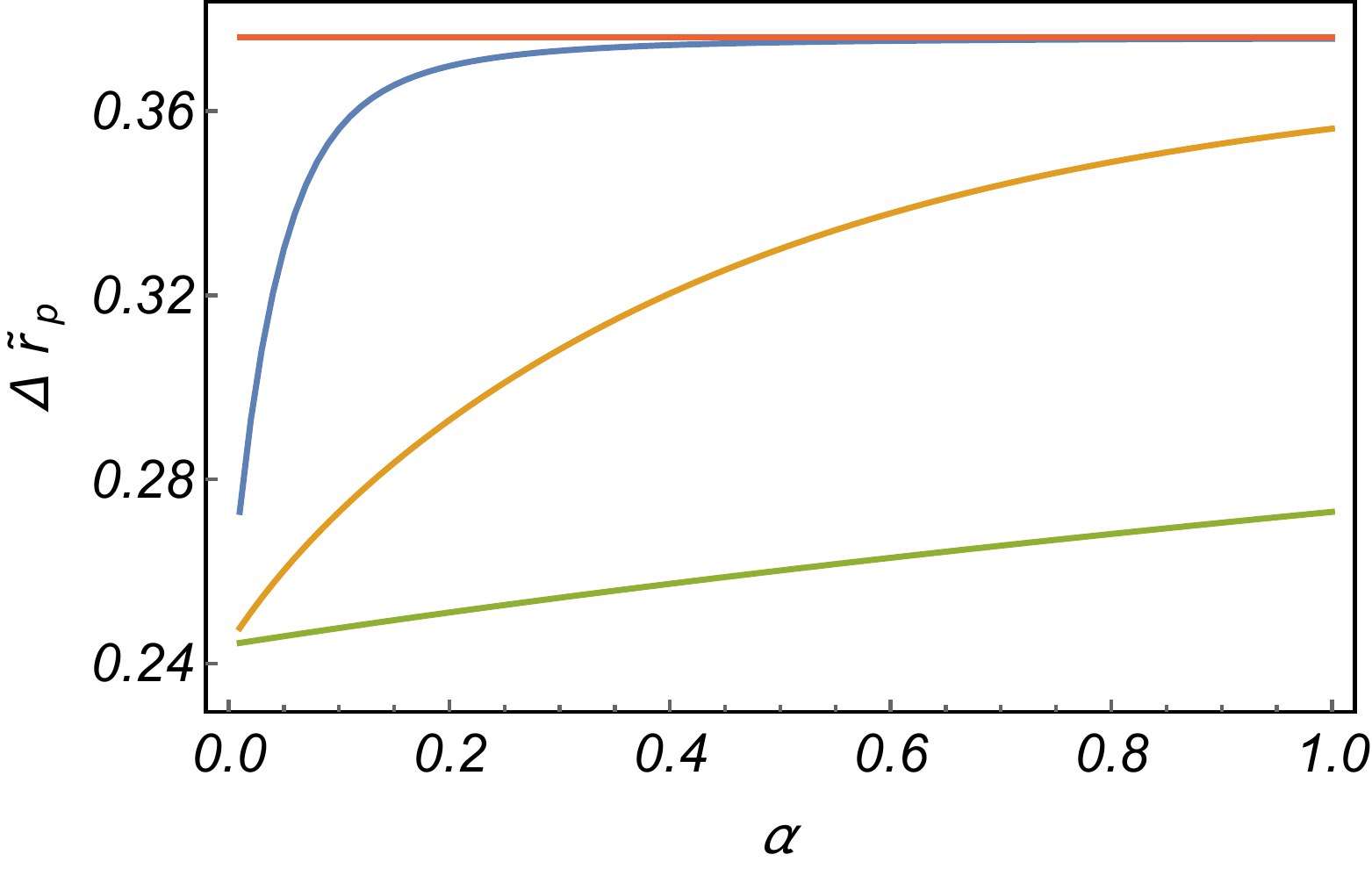}} 
	\caption{Behaviours of the {\red{reduced}} small/large photon sphere radius as a function of {\red{$\alpha$ for 5-dimensional Gauss-Bonnet AdS black holes with charge $Q = 0$ (red line), 0.01(blue line), 0.1(orange line) and 1(green line) at}} $\tilde{P}=0.98$ : (a) $\td{r}_{ps}$, (b) $\td{r}_{pl}$, (c) $\Delta \td{r}_{p}$.}
	\label{5a}
\end{figure*}

\begin{figure*}[htbp]
        \centering
        \subfigure[]{\label{6small}\includegraphics[width=5.9cm]{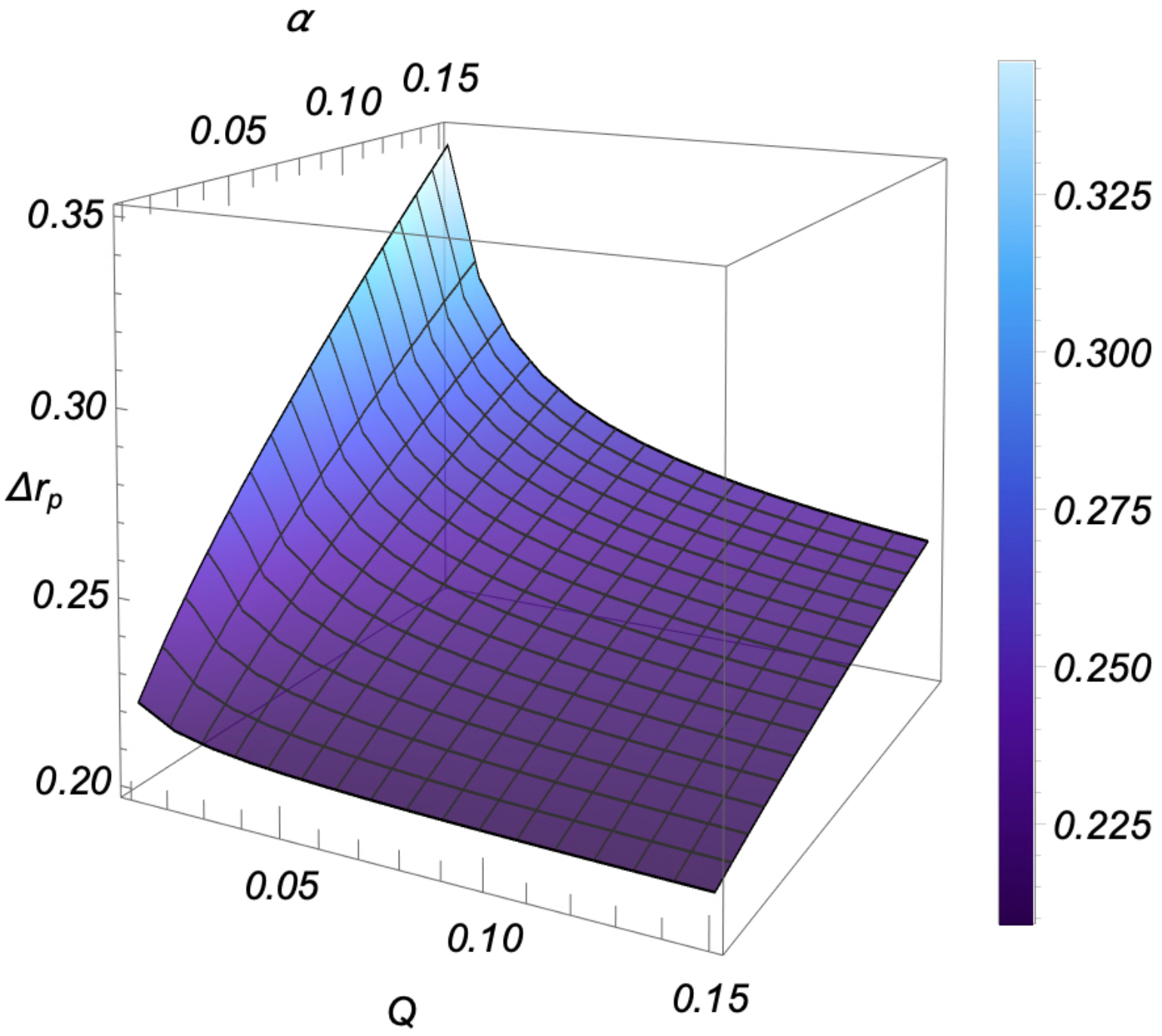}}
        \subfigure[]{\label{6large}\includegraphics[width=5.9cm]{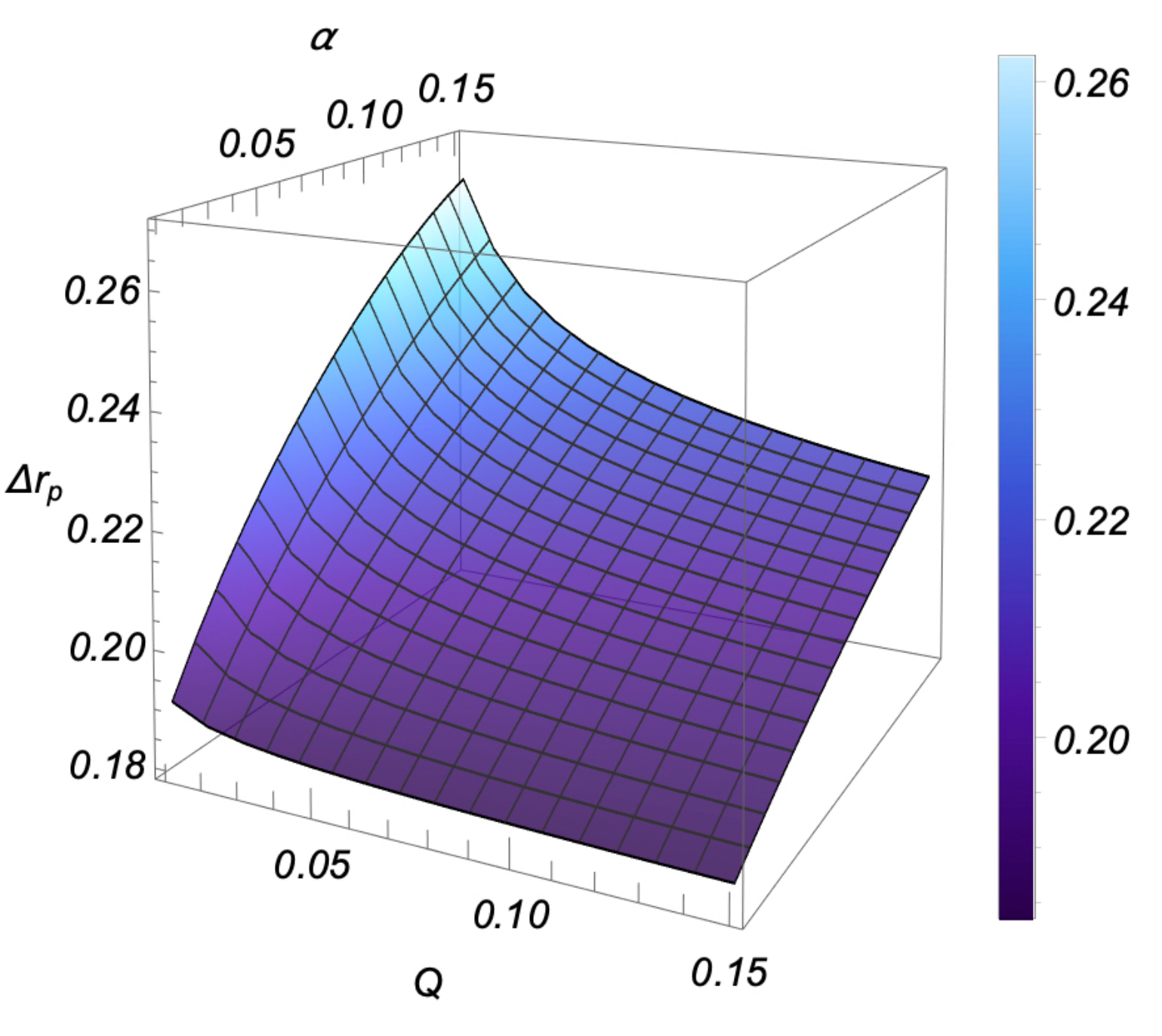}}
        \subfigure[]{\label{6large}\includegraphics[width=5.9cm]{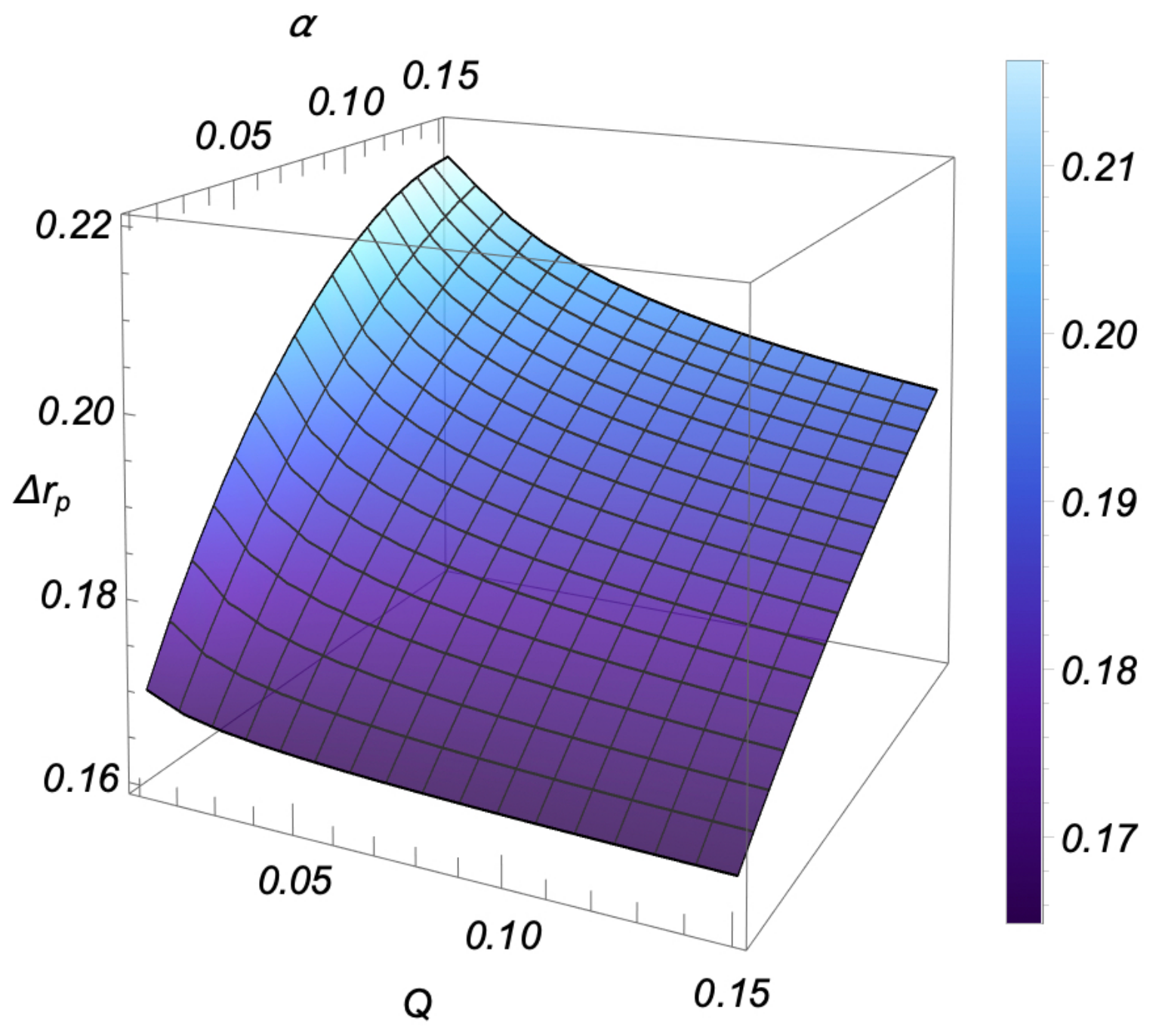}} \\
        \subfigure[]{\label{6large}\includegraphics[width=5.9cm]{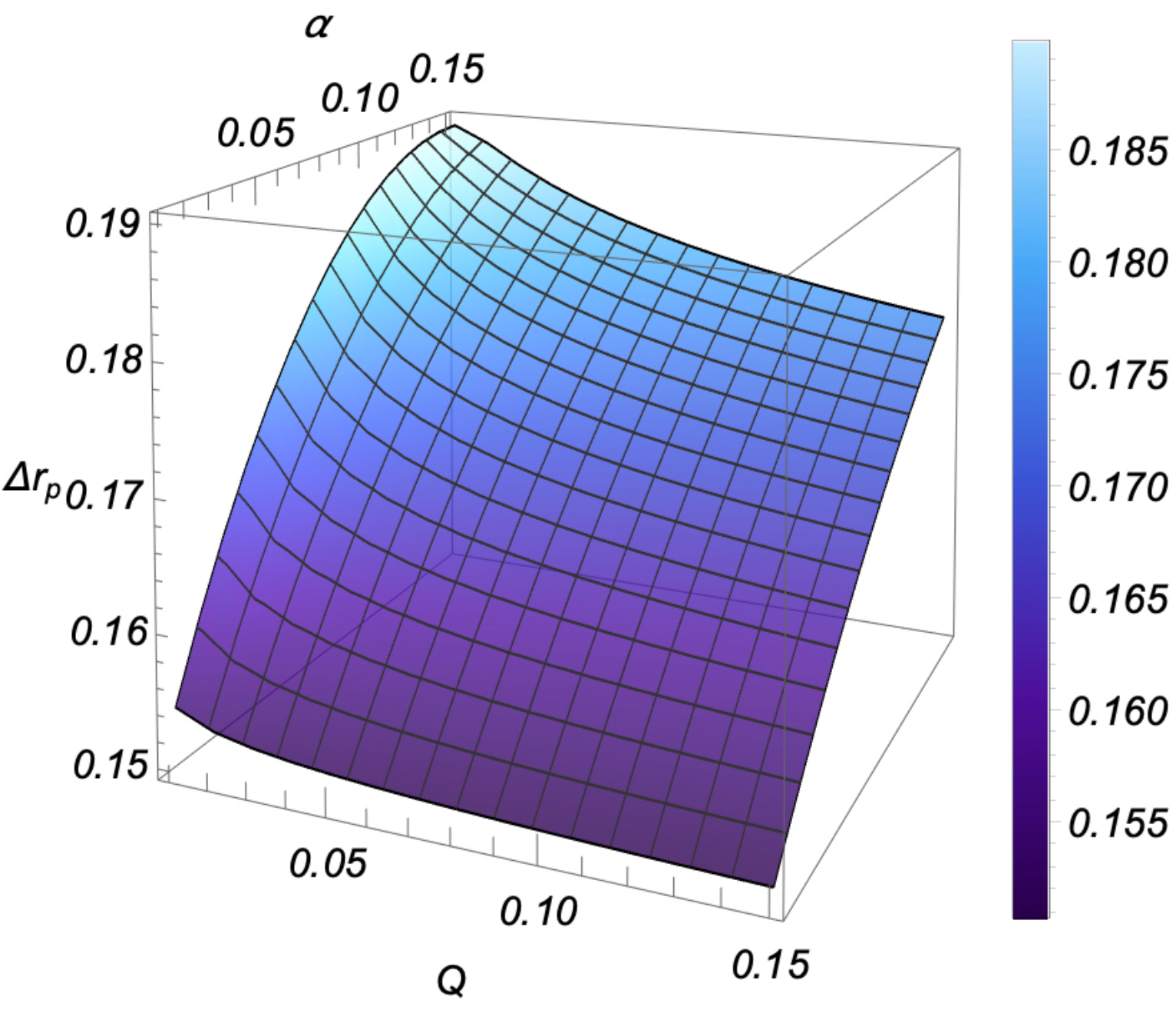}}  \    \   \     \
        \subfigure[]{\label{6large}\includegraphics[width=5.9cm]{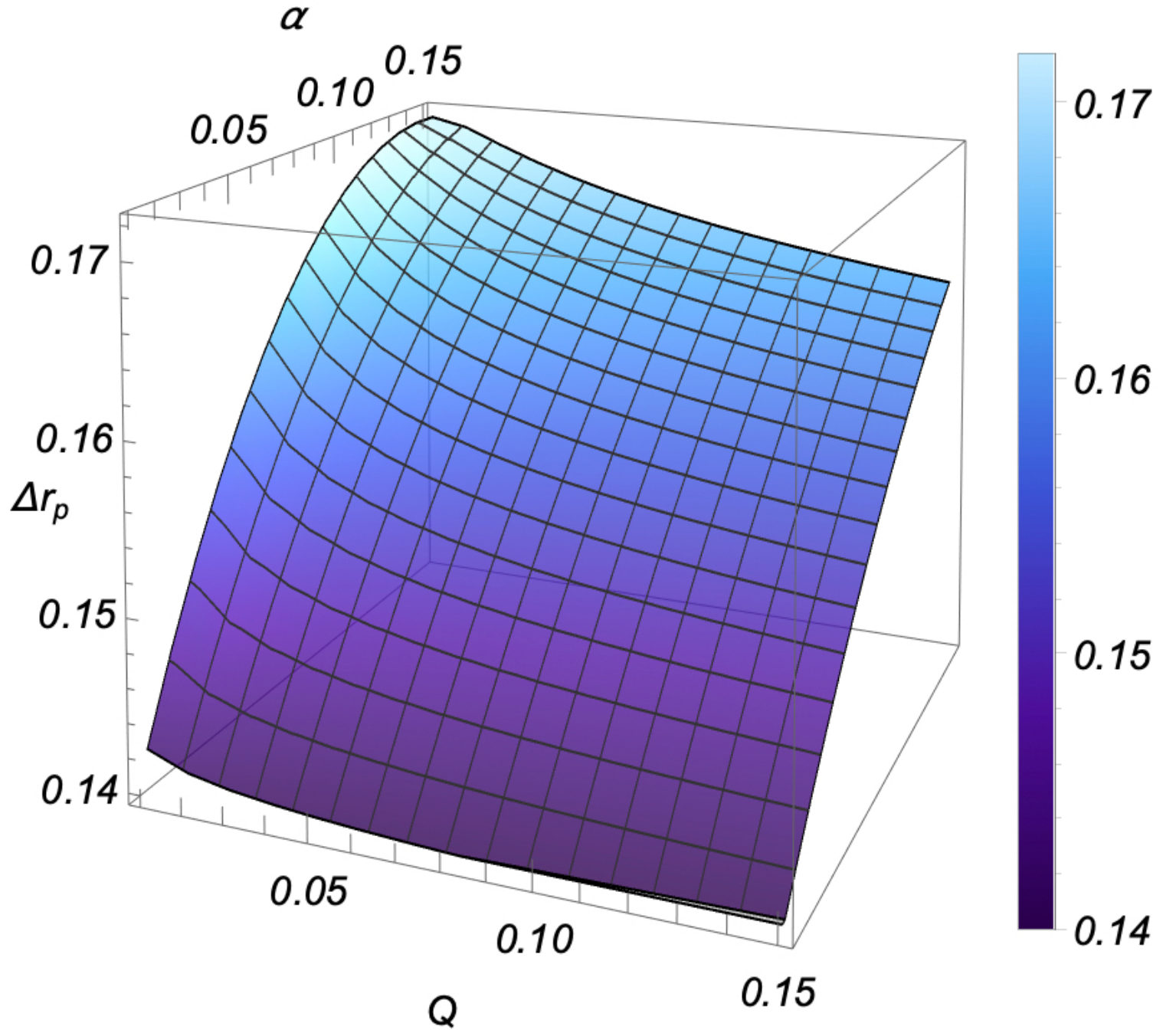}} 
        \caption{Behaviours of the {\red{reduced}} coexistence region as a function of $Q$ and $\alpha$ for {\red{$d$}}-dimensional charged Gauss-Bonnet AdS black holes with $\tilde{P}=0.98$ when (a ) $d=6$, (b) $d=7$, (c) $d=8$, (d)$d=9$, (e) $d=10$. Here, charge $Q$ and Gauss-Bonnet coefficient $\alpha$ range from 0.01 to 0.15, respectively.}
        \label{67}
\end{figure*}

According to Ref \cite{2018PhRvD..97j4027W}, we can find that {\red{reduced}} small/large photon sphere radius doesn't depend on the charge of {\red{$d$}}-dimensional RN-AdS black holes. In order to examine if it holds for {\red{$d$}}-dimensional charged GB-AdS black holes, we illustrate the {\red{impact}} of charge $Q$ on the {\red{reduced}} small/large photon sphere radius {\red{in \fig{5Q}}}. {\red{One can find, given a Gauss-Bonnet coefficient $\alpha$, reduced small photon sphere radius $\td{r}_{ps}$ increases with $Q$ shown in \fig{smallQ5}, while reduced large photon sphere radius $\td{r}_{pl}$ decreases with $Q$ shown in \fig{largeQ5}. Then we obtain the reduced coexistence region $\Delta \td{r}_{p}$, which decreases with $Q$ shown in \fig{coQ5}. Also, the impact of Gauss-Bonnet coefficient $\alpha$ is {\red{illustrated in \fig{5a}}}. One can easily find, given a charge value $Q$, reduced small photon sphere radius $\td{r}_{ps}$ decreases with $\alpha$ shown {\red{in  \fig{smalla5}}}, while reduced large photon sphere radius $\td{r}_{pl}$ increases with $\alpha$ shown in \fig{largea5}. Reduced coexistence region $\Delta \td{r}_{p}$ increases with $\alpha$ shown in \fig{coa5}. It's worth noting here that when charge vanishes, reduced small/large photon sphere radius and reduced coexistence region will not change with Gauss-Bonnet coefficient (red lines in \fig{5a}). In fact, according to Ref.\cite{2013JHEP...09..005C}, one knows that even when the charge of the black hole is absent, the small/large black hole phase transition still appears. It seems that the Gauss-Bonnet coefficient plays the same role as the charge. At this moment, we further show that, when the charge vanishes,  reduced small/large photon sphere radius doesn’t depend on the Gauss-Bonnet coefficient. It reminds us of the case of RN-AdS black holes. Apparently, Gauss-Bonnet coefficient is indeed acting the role as the charge of RN-AdS black holes.}}

\subsection{$d\ge6$ dimensional charged GB-AdS black holes}

\begin{figure}[htp]
	\centering
	\includegraphics[width=\columnwidth]{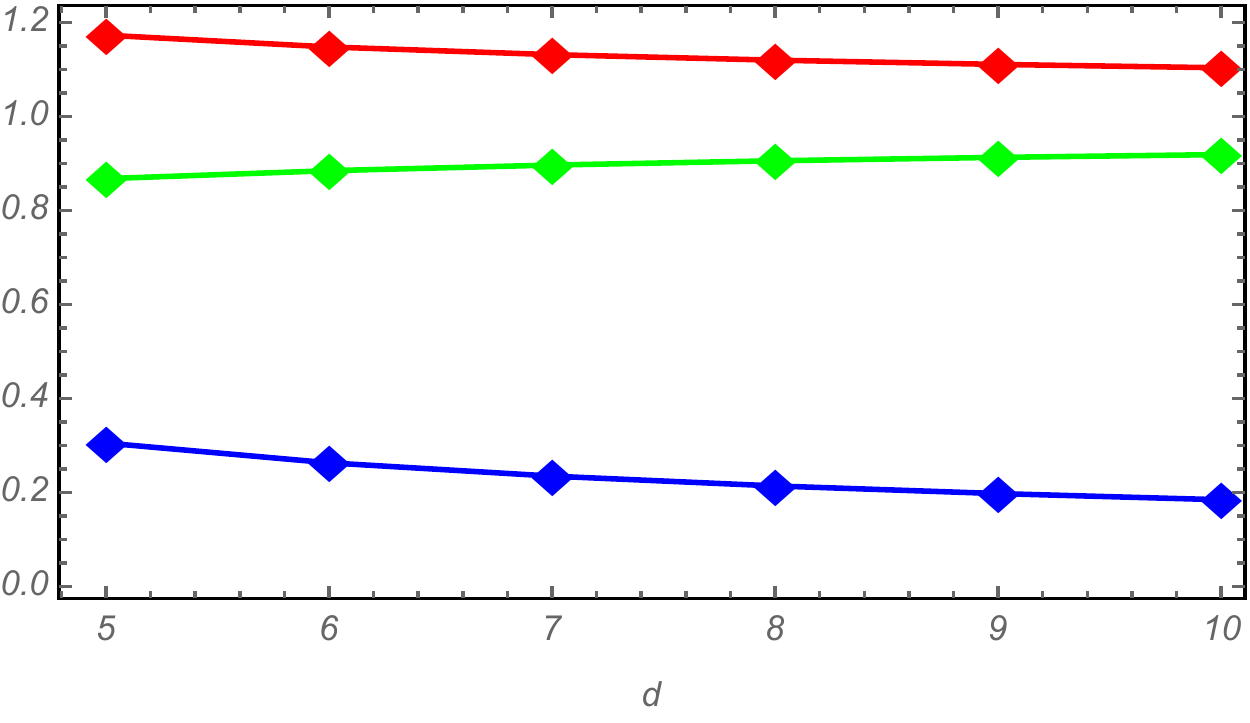}
	\caption{The change of {\red{reduced}} small/large photon sphere radius with dimensions. Green dot and red dot denote the small and large photon sphere radius, $r_{ps}$ and $r_{pl}$, respectively. Blue dot denotes the coexistence region $\Delta r_{p}$. Here, we set $Q=1, \alpha=0.1{,\red{\tilde{P}=0.97}}$.}
	\label{dimension}
\end{figure}

For higher dimensional charged GB-AdS black holes, {\red{we will study the cases of $d=6 -10$. In order to reduce the redundancy, the impact of $Q$ and $\alpha$ has been put into the same 3D diagram. As shown in}} \fig{67}, {one can see that} previous discussions are basically valid in the high-dimensional situation, except for the case when $Q\ll\alpha$. In this case, because no phase transition occurs, there is no so-called small/large photon sphere. In fact, just as concluded in Ref.\cite{2013JHEP...09..005C}, there is an upper bound on $\alpha/Q$ for SBH-LBH phase transition. {\red{The region where $Q\ll\alpha$ will not make any sense, so we only show the reduced}} coexistence region ranging from $0.1$ to $0.15$ for $Q$ and $0.1$ to $0.15$ for $\alpha$. In addition, setting $Q=1,\alpha=0.1, {\red{ \tilde{P}=0.97}}$, we plot \fig{dimension}, which shows that {\red{reduced}} small photon sphere radius $\td{r}_{ps}$ increases with dimension, while {\red{reduced}} large photon sphere radius $\td{r}_{pl}$ decreases with dimension for $d=5-10$. The {\red{reduced}} coexistence region {\red{$\Delta \td{r}_{p}$}} decreases with dimension. 

\section{Discussions and Conclusions}\label{sec5}

In this paper, we study the relationship between the photon sphere and black hole phase transition in the reduced parameter space. We show that, there exist nonmonotonic behaviors of photon sphere radius, not only for charged RN-AdS black holes,  but also for charged GB-AdS black holes. Also, we can use this nonmonotonic behaviors to characterize the SBH-LBH phase transition. What's more, using equal law in $P-V$ plane, we get the coexistence curve and further determine the boundary of coexistence region---saturated small photon sphere radius and saturated large photon sphere radius {\red{in the reduced parameter space. We find the reduced coexistence region decreases with charge $Q$ while increases with Gauss-Bonnet coefficient $\alpha$ for 5-dimensional charged-GB-AdS black holes.  Apparently, even though the quantities have been parameterized in the reduced space. The reduced coexistence region still depends on the charge $Q$ due to the presence of Gauss-Bonnet coefficient $\alpha$, that's different from the case of RN-AdS black holes\cite{2018PhRvD..97j4027W}. One more thing, different from the case of RN-AdS black holes, when charge is absent the small/large black hole phase transition still appears and so does the small/large photon sphere radius. What's more, we find reduced coexistence region doesn't change with Gauss Bonnet coefficient $\alpha$ at this moment. In this sense, the Gauss Bonnet coefficient $\alpha$ is acting the role as the charge of RN-AdS black holes. }}For the higher-dimensional charged GB-AdS black holes, we study the case of $d= 6-10$. The results show that previous conclusions still hold. Furthermore, we find that {\red{reduced}} coexistence region {\red{$\Delta \td{r}_{p}$}} decreases with dimension for $d=6-10$.

So far, we have shown how the charge, Gauss-Bonnet coefficient and dimension affect the coexistence region in the reduced parameter space for {\red{$d$}}-dimensional GB-AdS black holes and point out some observational differences between Einstein gravity (taking the RN-AdS black holes as an example) and Einstein-Gauss-Bonnet gravity with the help of the photon sphere. However, it's worth noting that we have not investigated the precise upper bound of $Q/\alpha$ that allows SBH-LBH phase transition to occur in the higher dimensional charged-AdS-GB black holes. More details about that, one can refer to Ref. \cite{2013JHEP...09..005C}.

\acknowledgements{This work is supported by the National Natural Science Foundation of China (Grant No.11235003). S. Han also thanks Overseas Study Fellowship Project from Physics Department of Beijing Normal University. And his research at Perimeter Institute is supported in part by the Government of Canada through the Department of Innovation, Science and Economic Development and by the Province of Ontario through the Ministry of Research, Innovation and Science.

%

\bibliography{refs}
\bibliographystyle{utphys}

\end{document}